\documentclass[11pt,epsf]{article}
\usepackage{amsmath}
\usepackage{amsfonts}
\usepackage{amssymb}
\usepackage{amsthm}
\usepackage{graphicx}
\usepackage{xcolor}
\usepackage{comment}

\newcommand{\rfec}{{r_{\textnormal{fec}}}}

\newtheorem{theorem}{Theorem}

\theoremstyle{definition}
\newtheorem{remark}{Remark}
\newcommand{\dfn}{\stackrel{\triangle}{=}}
\newcommand {\exe} {\stackrel{\cdot} {=}}
\newcommand {\lexe} {\stackrel{\cdot} {\le}}
\newcommand{\eqa}{\stackrel{\mbox{\tiny (a)}}{=}}
\newcommand{\eqb}{\stackrel{\mbox{\tiny (b)}}{=}}
\newcommand{\eqc}{\stackrel{\mbox{\tiny (c)}}{=}}

\newcommand{\gea}{\stackrel{\mbox{\tiny (a)}}{\ge}}
\newcommand{\geb}{\stackrel{\mbox{\tiny (b)}}{\ge}}

\newcommand {\reals} {{\rm I\!R}}
\newcommand {\ba} {\mbox{\boldmath $a$}}
\newcommand {\bb} {\mbox{\boldmath $b$}}
\newcommand {\bc} {\mbox{\boldmath $c$}}

\newcommand {\bv} {\mbox{\boldmath $v$}}
\newcommand {\bw} {\mbox{\boldmath $w$}}
\newcommand {\bx} {\mbox{\boldmath $x$}}
\newcommand {\by} {\mbox{\boldmath $y$}}

\newcommand {\bE} {\mbox{\boldmath $E$}}

\newcommand {\bG} {\mbox{\boldmath $G$}}

\newcommand {\hP} {\hat{P}}
\newcommand {\hH} {\hat{H}}
\newcommand {\hI} {\hat{I}}

\newcommand {\bX} {\mbox{\boldmath $X$}}
\newcommand {\bY} {\mbox{\boldmath $Y$}}

\newcommand {\tW} {\tilde{W}}
\newcommand {\tH} {\tilde{H}}

\newcommand{\calC}{{\cal C}}

\newcommand{\calE}{{\cal E}}

\newcommand{\calI}{{\cal I}}

\newcommand{\calM}{{\cal M}}

\newcommand{\calT}{{\cal T}}

\newcommand{\calX}{{\cal X}}
\newcommand{\calY}{{\cal Y}}

\allowdisplaybreaks

\begin{document}
\thispagestyle{empty}
\setcounter{page}{1}
\setlength{\baselineskip}{1.5\baselineskip}
\title{Codebook Mismatch Can Be Fully Compensated\\by Mismatched Decoding}
\author{Neri Merhav\thanks{N. Merhav is with the Viterbi Faculty of Electrical
and Computer Engineering, Technion -- Israel Institute of Technology, Haifa
3200003, Israel.
Email: {\tt merhav@technion.ac.il}}
\and
Georg B\"ocherer\thanks{G. B\"ocherer is with the Munich Research Center, Huawei Technologies Duesseldorf GmbH, Germany. Email: {\tt georg.boecherer@ieee.org} 
}} 
\maketitle

\begin{abstract}
We consider an ensemble of constant composition codes that are subsets of
linear codes: while the encoder uses only the constant-composition subcode,
the decoder operates as if the full linear code was used, with
the motivation of simultaneously benefiting both from the probabilistic
shaping of the channel input and from the linear structure of the code.
We prove that the codebook mismatch can be fully compensated by using a
mismatched additive decoding metric that achieves the random coding error
exponent of (non-linear) constant
composition codes. As the coding rate tends to the mutual information, the
optimal mismatched metric approaches the maximum a posteriori probability (MAP) metric, 
showing that codebook mismatch with mismatched MAP metric is
capacity-achieving for the optimal input assignment.\\

\end{abstract}
\clearpage
\section{Introduction}

As is very well known, linear codes have always been of central interest
in channel coding theory, thanks to their convenient practical implementation,
both at the encoder and the decoder side (see, e.g., \cite[Chap.\ 6]{Gallager68},
\cite[Part II]{McEliece77}, \cite[Sections 2.9, 2.10]{VO79} for major
elementary textbooks, as well as a vast amount of other books and articles). 
The structure of linear codes, together with the additivity of the optimal
channel decoding metric of certain memoryless channels, can
offer reduced decoding complexity in many ways, such as in syndrome
decoding (for relatively high coding rates), bounded distance decoding \cite[Sect.\ 6.2]{LC04}, Chase decoding
\cite{Chase72}, and many other decoding schemes that are based on temporary hard
decision, followed by a process
of correction using soft information. In many cases, the resulting decoding is
equivalent (or at least asymptotically so) to the optimal maximum likelihood
(ML) decoding. 
Also, for
convolutional codes, which form a special subclass of linear codes, 
the Viterbi decoder offers dramatic reduction in
computational complexity \cite{VO79} 
without sacrificing decoder optimality. Last but not least, polar codes, invented
by Stolte~\cite{stolte2002rekursive} and Ar{\i}kan~\cite{Arikan09} and proven
to be capacity-achieving by Ar{\i}kan~\cite{Arikan09}, form another subclass of linear codes.
Polar codes are perhaps the most attractive codes in the front line of research in coding theory today, with decoding complexity that is proportional to $n\log n$, where $n$ is the block
length.\footnote{There are, of course, additional classes of modern codes with
efficient decoding schemes, like Turbo codes and low-density parity check
codes, but their decodings are iterative and so, in general they are not
guaranteed to be equivalent to ML decoding.}
In general, it would be safe to say that most of the
modern practical codes are linear. For channels with a sufficient degree of
symmetry, like binary-input, output-symmetric (BIOS) channels, 
whose capacity-achieving input distribution is uniform, it is well
known that linear codes can achieve capacity, as linear codes inherently
induce the uniform input distribution. Moreover, the ensemble of linear codes
achieves the well known random coding error exponent at all coding rates up to
channel capacity \cite[Theorem~6.2.1]{Gallager68}.

An inherent limitation of linear codes, however, is that they cannot achieve
capacity (and hence neither can they achieve the channel's reliability function) 
when the capacity-achieving input distribution is non-uniform. 
One way to compensate for this drawback, and to achieve capacity
with linear codes nonetheless, is to
extend the channel input alphabet using a many-to-one mapping that induces
the desired input distribution (which is an operation also known as
``probabilistic shaping''), and use the
linear code on symbols of the extended alphabet -- see, e.g.,
Gallager \cite[p.~208]{Gallager68} for the details.
This approach is conceptually simple, however, not very attractive from the practical point of
view, because the required extended alphabet is often much larger, and so, many
more coded bits need to be processed, and the inevitable consequence is
increased complexity and increased power consumption.

During the years, researchers in coding theory have been pondering about the
question whether it is possible to enjoy the best of both words, namely,
to achieve capacity (as well as good error-rate performance at lower rates) without
giving up on the above-mentioned benefits of the linear code structure and the additive metric
decoding, and without paying the price of increased complexity of
Gallager's alphabet
extension needed for probabilistic shaping.

In this context, the idea of probabilistic amplitude shaping (PAS)
\cite{BSS15} was
proposed to have exactly the above-mentioned features: it enables the use of
linear codes with non-uniform distribution without the need of alphabet
extension. The basic idea is as follows: at the transmitter, the uniformly
distributed message
bits are transformed into ``probabilistically shaped'' codewords with the desired distribution. This
can be achieved efficiently using
distribution matching (DM) algorithms~\cite{schulte2020algorithms}. The shaped sequences are then encoded
systematically with a linear code.
The systematic encoding preserves the imposed distribution of the message
part, and generates in addition uniformly distributed parity bits. For
input distributions that have a uniform distribution as a factor, the parity
bits can be used to generate the uniform
factor, and the shaped bits can be used for the non-uniform factor. For the
additive white Gaussian noise (AWGN) channel with $M$-ary amplitude shift keying
(ASK) input alphabet $\{\pm 1, \pm 3, \ldots, \pm(M-1)\}$, the uniform factor
is the distribution of the symbol signs, and
the non-uniform factor is the distribution of the symbol amplitudes, hence the
name probabilistic amplitude shaping.
In \cite{BSS19}, the PAS encoding was extended to linear layered probabilistic shaping (LLPS), which also allows for shaped parity bits.
The decoding rule for PAS~\cite{BSS19} and LLPS~\cite{BSS15} is essentially the maximum a posteriori (MAP) decoding rule, which takes into account the input shaping.
The schemes proposed in
\cite{BSS19}, \cite{BSS15} suggest that in
principle, linear
codes can be used for channels with non-uniform input distribution, by
encoding into those codewords that have the required distribution.

For
instance, suppose that for a channel with input distribution $P_X$, 
only codewords of type $P_X$ are used. Thus, the set of codewords that are actually transmitted
forms a constant composition code~\cite[Chapter~10]{CK11}, yet the decoder
decodes as it would even if the entire (linear) codebook was used. Therefore,
one can think of this structure
as a linear extension of the constant composition code.
This introduces a {\em codebook mismatch} between the set of
transmitted codewords (e.g., a constant composition code) and the set of
hypotheses for decoding (e.g., the linear code).

Several previous works have discussed codebook mismatch, linear codes, and additive metrics for analyzing probabilistic shaping schemes. 
In \cite[Section~I]{achtenberg2013theoretic}, codebook mismatch was discussed and judged to be suboptimal. 
The works
\cite{bocherer2016achievable},
\cite[Chapter~7]{bocherer2018principles},
\cite{BSS19}, and
\cite{bocherer2020probabilistic}, 
model codebook mismatch by considering a random code generated according to a 
uniform distribution as the extended codebook used for decoding. In \cite{bocherer2016achievable}, 
a discrete memoryless channel (DMC) is considered and a typicality decoder is used. 
Discrete-input, continuous-output channels and additive decoding metrics are considered in 
\cite{bocherer2018principles},
\cite{BSS19}, 
\cite{bocherer2020probabilistic},
and achievable rates for successful encoding and successful decoding are analyzed separately. The encoding error analysis by typicality in \cite[Section~7.3]{bocherer2018principles} is simplified in \cite[Appendix~A]{BSS19} by using a simple counting argument. In \cite{bocherer2020probabilistic}, encoding and decoding rates are combined to an achievable rate without further proof. The work \cite[Chapter~10]{bocherer2018principles} 
derives an error exponent for PAS and shows that PAS using constant composition DM (CCDM)~\cite{schulte2016constant} and MAP 
decoding on a random linear extension code achieves the mutual information of discrete-input, 
continuous-output memoryless channels. 
This result implies that PAS achieves capacity for a number of practically relevant channels, including the AWGN channel with ASK input. Note that the related works \cite{amjad2018information} and \cite{gultekin2020achievable} do not model the codebook mismatch in their analysis. In \cite{amjad2018information}, a joint source channel coding scenario is considered, while in \cite{gultekin2020achievable}, decoding is performed on the set of shaped sequences and no extension code is considered for decoding.

It is this background that motivates the study of codebook mismatch with linear extension codes and additive decoding metrics 
that we conduct in this
work. Instead of considering the PAS configuration, we examine the 
following, simpler, and more general setup, whose focus is on harnessing linear codes for constant-composition coding: 
Consider the ensemble of linear codes, where
the encoder uses only a subset of the codebook that forms a
constant-composition code (corresponding to a certain type class), 
whereas the decoder uses an arbitrary additive decoding metric (e.g., the ML or MAP decoding
metric) and decodes the same way as if
the full linear code was used, ignoring the fact that non-typical codewords
are actually never used by the encoder. The motivation for the decoder not
to discard the non-typical codewords is in order to maintain the linear structure of
the code, along with its benefits, as described earlier.
The main questions that we concern ourselves
with are the following: 
\begin{enumerate}
\item For a given discrete memoryless channel (DMC) and a given decoding
metric, what is the random coding error exponent
associated with this setting? 
\item Considering the fact that, due to the
codebook mismatch, the decoder is
sub-optimal, can we choose an alternative additive decoding metric that would
compensate for the codebook mismatch?
\end{enumerate}
In response to these two questions, we first derive the exact error exponent for linear
codes under code mismatch for a given, additive decoding metric, and then
show that by optimizing this decoding metric, we can improve the random
coding exponent so as to coincide with that
of general (non-linear) constant composition codes
\cite[Theorem~10.2]{CK11}, which means, among other things, that capacity is
achieved. More specifically, regarding item no.\ 1 above, we show that the
error exponent of any given additive metric cannot be larger than the random
coding exponent of the ensemble of fixed composition codes, but on the other
hand, with regard to item no.\ 2, we fully characterize the optimal metric that achieves this upper
bound. We further show that as rate approaches mutual information, the optimal metric 
becomes the MAP metric, implying that MAP decoding 
on the linear extension code achieves capacity, given that the constant 
composition is the capacity-achieving input distribution.

The remainder of this work is organized as follows. 
In Section \ref{notation}, we establish the notation conventions.
In Section \ref{ps}, we formalize the problem setting and spell out the objectives of this work.
In Section \ref{mainresult}, we present the main theorems and discuss them.
Finally, in Section \ref{proof}, we prove the theorems.

\section{Notation}
\label{notation}

Throughout the paper, random variables will be denoted by capital
letters, specific values they may take will be denoted by the
corresponding lower case letters, and their alphabets
will be denoted by calligraphic letters. Random
vectors and their realizations will be denoted,
respectively, by capital letters and the corresponding lower case letters,
both in the bold face font. Their alphabets will be superscripted by their
dimensions. For example, the random vector $\bX=(X_1,\ldots,X_n)$, ($n$ --
positive integer) may take a specific vector value $\bx=(x_1,\ldots,x_n)$
in $\calX^n$, the $n$--th order Cartesian power of $\calX$, which is
the alphabet of each component of this vector.
Sources and channels will be denoted by the letter $P$, $Q$, and $W$,
sometimes subscripted by the names of the relevant random variables/vectors and their
conditionings, if applicable, following the standard notation conventions,
e.g., $P_X$, $Q_{Y|X}$, and so on. When there is no room for ambiguity, these
subscripts will be omitted.
The probability of an event $\calE$ will be denoted by $\mbox{Pr}\{\calE\}$,
and the expectation
operator will be
denoted by $\bE\{\cdot\}$.
For two positive sequences $a_n$ and $b_n$, the notation $a_n\exe b_n$ will
stand for equality in the exponential scale, that is,
$\lim_{n\to\infty}\frac{1}{n}\log \frac{a_n}{b_n}=0$. Similarly,
$a_n\lexe b_n$ means that
$\limsup_{n\to\infty}\frac{1}{n}\log \frac{a_n}{b_n}\le 0$, and so on.
The indicator function
of an event $\calE$ will be denoted by $\calI\{E\}$. The notation $[x]_+$
will stand for $\max\{0,x\}$. Logarithms will be defined to the base 2, unless
specified otherwise.

The empirical distribution of a sequence $\bx\in\calX^n$, which will be
denoted by $\hat{P}_{\bx}$, is the vector of relative frequencies
$\hat{P}_{\bx}(x)$
of each symbol $x\in\calX$ in $\bx$.
The type class of $\bx\in\calX^n$, denoted $\calT(\bx)$,
is the set of all
vectors $\bx'$
with $\hat{P}_{\bx'}=\hat{P}_{\bx}$. 
When we wish to emphasize the
dependence of the type class on the empirical distribution, say $P$, we
will denote it by
$\calT(P)$. Information measures associated with empirical distributions
will be denoted with `hats' and will be subscripted by the sequences from
which they are induced. For example, the entropy associated with
$\hat{P}_{\bx}$, which is the empirical entropy of $\bx$, will be denoted by
$\hat{H}_{\bx}(X)$. An alternative notation, following the conventions
described in the previous paragraph, is $H(\hP_{\bx})$.
Similar conventions will apply to the joint empirical
distribution, the joint type class, the conditional empirical distributions
and the conditional type classes associated with pairs (and multiples) of
sequences of length $n$.
Accordingly, $\hP_{\bx\by}$ would be the joint empirical
distribution of $(\bx,\by)=\{(x_i,y_i)\}_{i=1}^n$,
$\calT(\bx,\by)$ or $\calT(\hP_{\bx\by})$ will denote
the joint type class of $(\bx,\by)$, $\calT(\bx|\by)$ will stand for
the conditional type class of $\bx$ given
$\by$, $\hH_{\bx\by}(X,Y)$ will designate the empirical joint entropy of $\bx$
and $\by$,
$\hH_{\bx\by}(X|Y)$ will be the empirical conditional entropy,
$\hI_{\bx\by}(X;Y)$ will
denote empirical mutual information, and so on. 

Given a fixed probability
assignment, $P_X$, of a random variable $X$, and given a generic conditional
distribution $Q_{Y|X}$, we denote the induced information measures using the
conventional notation rules of the information theory literature, but with the
subscript $Q$. For example, $I_Q(X;Y)$, $H_Q(Y)$,
$H_Q(Y|X)$, and  $H_Q(X|Y)$ will denote, 
respectively, the mutual information between $X$ and $Y$,
the marginal entropy of $Y$, the conditional entropy of $Y$ given $X$ and the
conditional entropy of $X$ given $Y$, all induced by $P_X\times Q_{Y|X}$. 
Likewise, given $P_X$ and $Q_{Y|X}$, the induced conditional distribution of
$X$ given $Y$ will be denoted by $Q_{X|Y}$. The
same notation conventions will apply whenever other auxiliary random variables
will be involved, such as $X'\sim P_X$. The weighted Kullback-Leibler divergence
between two conditional distributions, say, $Q_{Y|X}$ and
$W=\{W(y|x),~x\in\calX,~y\in\calY\}$, is defined as
\begin{equation}
D(Q_{Y|X}\|W|P_X)=\sum_{x\in\calX}P_X(x)\sum_{y\in\calY}Q_{Y|X}(y|x)\log\frac{Q_{Y|X}(y|x)}{W(y|x)}.
\end{equation}

\section{Problem Setting}
\label{ps}

We consider coded communication via a discrete memoryless channel (DMC) 
with a finite input alphabet $\calX$, a finite output alphabet
$\calY$, and a single--letter transition probability matrix,
$W=\{W(y|x),~x\in\calX,~y\in\calY\}$. When the channel is fed by an input
vector $\bX=\bx=(x_1,\ldots,x_n)\in\calX^n$, it outputs a random vector
$\bY=(Y_1,\ldots,Y_n)\in\calY^n$, according to the conditional probability
distribution,
\begin{equation}
\mbox{Pr}\{\bY=\by|\bX=\bx\}=W(\by|\bx)=\prod_{i=1}^nW(y_i|x_i).
\end{equation}
Without essential loss of generality, we assume the cardinality of
$\calX$ to be a power of two, i.e., $|\calX|=2^m$ for some positive integer $m$.
In the absence of this property,
one can always formally extend $\calX$ to be of the size of
$\exp_2\left(\lceil\log_2|\calX|\rceil\right)$, by adding to $\calX$ some 
dummy input symbols that are never actually used.
We adopt this assumption in order to allow the
restriction to binary linear codes, and thereby simplify the notation and the
derivations.

The transmitter is assumed to employ a
binary linear code of
block length $nm$ and code dimension $k=nm\rfec$, where
$0<\rfec\leq 1$. 
\begin{remark}
For $\rfec$, the subscript FEC stands for forward error correction and emphasizes following \cite{BSS19} that $\rfec$ is the code rate of the employed binary linear FEC code. As we will see below, the effective rate $R$ at which information is transmitted over the channel also depends on the type $P_X$ of the constant composition, i.e., it is not determined by the FEC code rate $\rfec$ alone.
\end{remark}
The encoding mechanism is as follows.
An information message, $w\in\{0,1,2,\dotsc,2^k-1\}$, with a binary representation
denoted by $\bb(w)\in\{0,1\}^k$, is
mapped into a codeword $\bc(w)$ according to
\begin{align}
\label{eq:linear code}
\bc(w) = \bb(w)\cdot G + \bv,
\end{align}
where $\bG\in\{0,1\}^{k\times nm}$, $\bv\in\{0,1\}^{nm}$,
and where the entries of $\bG$ and $\bv$ are selected independently at
random according to the
uniform distribution over $\{0,1\}$.
The binary codeword $\bc(w)$ is mapped into a channel input
vector $\bx(w)\in\calX^n$ via a labeling function $\phi\colon
\{0,1\}^m\to\calX$ that
indexes the symbols in the channel alphabet $\calX$ by $m$ bits, i.e.,
\begin{align}
\bc(w)\to \bx(w) = \phi(c_1(w)\dots c_m(w))\phi(c_{m+1}(w)\dots
c_{2m}(w))\dotsc\phi( c_{(n-1)m+1}(w)\dotsc c_{nm}(w)),
\end{align}
$c_i(w)$, $i=0,\ldots,nm-1$, being the components of $\bc(w)$.
We denote the codebook
\begin{equation}
\calC=\{\bx(w),~w\in\{0,1,\ldots,2^k-1\}\}.
\end{equation}

In contrast to the traditional setting, where all codewords of the linear code
are used, in this work, we consider the case where only a subset of the
codebook is used, namely, codewords, $\{\bx(w)\}$, which belong to a given type
class, $\calT(P_X)$, where $P_X$ is a certain empirical distribution over $\calX$. Since
$|\calT(P_X)|\exe 2^{nH(P_X)}$, and since the code partitions the Hamming
space, $\{0,1\}^{nm}$, into $2^{nm(1-\rfec)}$
disjoint cosets, each of size $2^{nm\rfec}$, then there must be at least one
coset that includes at least
$$\frac{|\calT(P_X)|}{2^{nm(1-\rfec)}}\exe
\frac{2^{nH(P_X)}}{2^{nm(1-\rfec)}}=2^{n[H(P_X)-m(1-\rfec)]}$$
codewords in $\calT(P_x)$. Our encoder will use (a subset of) such a coset, henceforth
denoted $\calC'$, to encode
information at the rate,
\begin{equation}
	\label{rates}
R=H(P_X)-m(1-\rfec),
\end{equation}
where we keep in mind the well known fact that the probability of error does
not depend on the coset representative. Thus, the effective coding rate, $R$,
associated with
$\calC'$, is always less than or equal to $m\rfec$, with equality iff $P_X$ is the
uniform distribution over $\calX$.

At the receiver side, a metric decoder is used. The decoding metric
is a function $U:\calX\times\calY\to\reals^+$, which induces the following
decoding rule:
\begin{align}
	\hat{w}(\by)=\mbox{arg max}_{w\in\{0,1,\dotsc,2^k-1\}}
U(\bx(w),\by),
\end{align}
where
\begin{align}
U(\bx(w),\by)=\prod_{i=1}^n U(x_i(w), y_i),
\end{align}
$x_i(w)$, $i=1,\ldots,n$, being the components of $\bx(w)$.
Note that for practical reasons (discussed in the Introduction), 
the decoder examines {\em all codewords of the original linear
code, $\calC$}, not only those of the subcode, $\calC'$, of $P_X$-typical codewords.
In other words, $\bx(\hat{w}(\by))$ can be any codeword in $\calC$, not
necessarily in $\calC'$.

The probability of error for a given $\bw$ (with $\bx(w)\in\calC'$) and a given code $\calC$,
is defined as
\begin{equation}
P_{\mbox{\tiny e}|w}(\calC)=\mbox{Pr}\{\hat{w}(\bY)\ne w\},
\end{equation}
where the randomness of $\bY$ is due to the channel only.
The average error probability is defined as
\begin{equation}
\bar{P}_{\mbox{\tiny
e}}=\bE\left\{\frac{1}{2^{nR}}\sum_{\{w:~\bx(w)\in\calC'\}}\bar{P}_{\mbox{\tiny
e}|w}(\calC)\right\},
\end{equation}
where the expectation is with respect to the randomness of $(\bG,\bv)$.
The random coding error exponent is defined as
\begin{equation}
E_{\mbox{\tiny r}}(R)=\lim_{n\to\infty}\left\{-\frac{\log
\bar{P}_{\mbox{\tiny e}}}{n}\right\},
\end{equation}
provided that the limit exists. In the sequel, whenever we need to emphasize the dependence
of the random coding error exponent upon the decoding metric, $U$, we will denote
it by $E_{\mbox{\tiny r}}(R,U)$.

Observe that our setting exhibits a situation of {\em codebook mismatch}: While
the encoder uses only the subcode, $\calC'$, the decoder acts as if the entire
larger code, $\calC$, was fully used, without taking advantage of the knowledge
that $\bx(w)$ must be in $\calC'$. This is done in order to avoid ruining the
linear structure of the code,
which is useful for fast decoding. Consequently, the decoder is suboptimal
even if its decoding metric is the maximum likelihood metric,
$U(x,y)=U_{\mbox{\tiny ML}}(x,y)\dfn W(y|x)$. The question that we study, in
this work, is whether $U_{\mbox{\tiny ML}}(x,y)$ can be replaced by another
decoding metric, $U(x,y)$, that would 
compensate for the codebook mismatch. 

Our main result in this work is in answering this question affirmatively. To
this end, we first derive a single--letter formula for $E_{\mbox{\tiny r}}(R,U)$,
for a given, arbitrary decoding metric, $U$, and then we demonstrate that
by maximizing $E_{\mbox{\tiny r}}(R,U)$ w.r.t.\ $U$, we can significantly improve
over $E_{\mbox{\tiny r}}(R,U_{\mbox{\tiny ML}})$ as well as over
$E_{\mbox{\tiny r}}(R,U_{\mbox{\tiny MAP}})$, where $U_{\mbox{\tiny
MAP}}(x,y)\dfn P_X(x)W(y|x)$. The comparison to $U_{\mbox{\tiny MAP}}$ is
relevant because, in spite of
the mismatch, it still achieves coding rates arbitrarily close to
the `capacity' associated with $P_X$, namely, the mutual information induced
by $P_X\times W$. Moreover, our main result is in proving that, on the one
hand, $E_{\mbox{\tiny
r}}(R,U)$ cannot exceed the random coding exponent of (non-linear) fixed
composition codes
\cite[Theorem 10.2]{CK11}:
\begin{equation}
\label{CKexponent}
E_{\mbox{\tiny
r}}^{\mbox{\tiny cc}}(R)=\min_{Q_{Y|X}}\left\{D(Q_{Y|X}\|W|P_X)+[I_Q(X;Y)-R]_+\right\}.\end{equation}
but on the other hand, we characterize the optimal decoding
metric, $U_*$, and show that it achieves $E_{\mbox{\tiny
r}}^{\mbox{\tiny cc}}(R)$.

\section{Main Results}
\label{mainresult}

Our main result is in the following theorem, whose proof can be found in
Appendix A.

\begin{theorem}
Consider the setting formulated in Section \ref{ps}.
\begin{enumerate}
\item For a given decoding metric $U$,
\begin{equation}
E_{\mbox{\tiny r}}(R,U)=\max_{0\le\rho\le 1}\sup_{\theta\ge 0}
\left\{-\sum_{x\in\calX}P_X(x)\log\left(\sum_{y\in\calY}\frac{W(y|x)}{[U_\theta(x|y)]^\rho}\right)
+\rho[H(P_X)-R]\right\},\label{eq:main result error exponent}
\end{equation}
where
\begin{equation}
U_\theta(x|y)\dfn\frac{[U(x,y)]^\theta}{\sum_{x'\in\calX}[U(x',y)]^\theta}.\label{eq:main result theta metric}
\end{equation}
\item For every metric $U$, $E_{\mbox{\tiny r}}(R,U)\le E_{\mbox{\tiny
r}}^{\mbox{\tiny cc}}(R)$.
\item For every given $\rho\in[0,1]$, 
assume that there exists a vector $Z=Z_\rho\dfn\{Z_\rho(x),~x\in\calX\}$, with strictly
positive components, that satisfies the system of simultaneous equations,
\begin{equation}
Z_\rho(x)=\sum_y[W(y|x)]^{1/(1+\rho)}\left[\sum_{x'}
\frac{P_X(x')[W(y|x')]^{1/(1+\rho)}}{Z(x')}\right]^\rho,~~~~\forall~x\in\calX,
\end{equation}
and define
\begin{equation}
U(x|y,\rho)=\frac{P_X(x)[W(y|x]^{1/(1+\rho)}/Z_\rho(x)}
{\sum_{x'}P_X(x')[W(y|x')]^{1/(1+\rho)}/Z_\rho(x')},
\end{equation}
and
\begin{equation}
U_\theta(x|y,\rho)=\frac{[U(x|y,\rho)]^\theta}{\sum_{x'}[U(x'|y,\rho)]^\theta}.
\end{equation}
Finally, let
\begin{equation}
	\rho_\star=\mbox{arg max}_{\rho\in[0,1]}\sup_{\theta\ge 0}
\left\{-\sum_{x\in\calX}P_X(x)\log\left(\sum_{y\in\calY}\frac{W(y|x)}{[U_\theta(x|y,\rho)]^\rho}\right)
+\rho[H(P_X)-R]\right\}.
\end{equation}
Then, the metric $U_\star\dfn \{U(x|y,\rho_\star),~x\in\calX,~y\in\calY\}$
achieves $E_{\mbox{\tiny
r}}^{\mbox{\tiny cc}}(R)$.
\end{enumerate}
\end{theorem}

\begin{remark}
The expression of $E_{\mbox{\tiny r}}(R,U)$ does not seem to lend itself to
closed form derivation of $U^*$ using traditional optimization techniques, and therefore, 
the proof of the third part of the theorem will be based on a chain of
inequalities relating $E_{\mbox{\tiny r}}(R,U)$ and $E_{\mbox{\tiny
r}}^{\mbox{\tiny cc}}(R)$ and examining the conditions under which the
inequalities become equalities.
As can be seen, the optimal metric, $U_\star$, that maximizes
$E_{\mbox{\tiny r}}(R,U)$, depends, in quite a complicated manner,
not only on the input assignment, $P_X$, and 
the channel, $W$, but also on the coding rate, $R$, via $\rho^*$.
\end{remark}
The remaining part of this section is devoted to a discussion on Theorem 1.
\subsection{Numerical Example of Error Exponents}

\begin{figure}[t]
\centering
\includegraphics{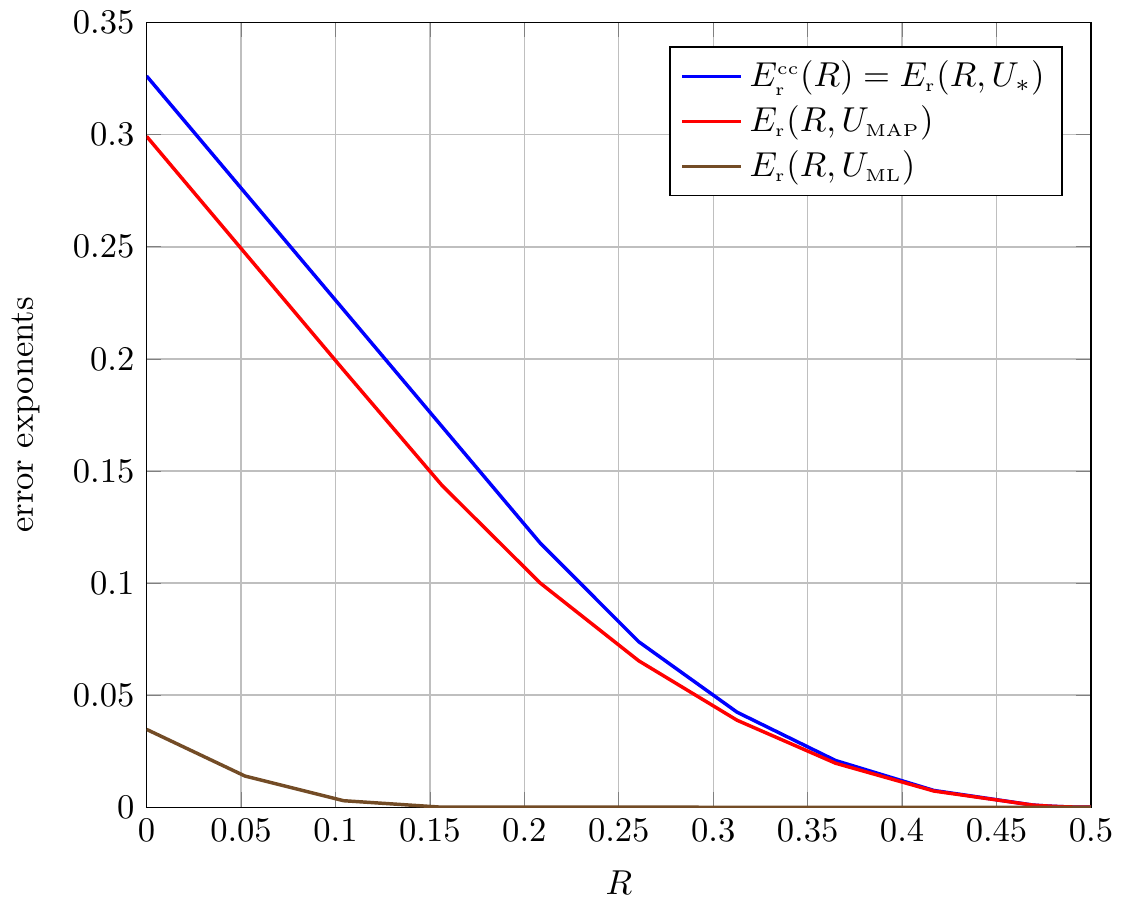}
\caption{Comparison of the random coding exponents associated with
the ML metric, $U_{\mbox{\tiny ML}}$,
the MAP metric, $U_{\mbox{\tiny MAP}}$, and the
random coding exponent of the ensemble of non-linear constant composition codes
(\ref{CKexponent}), which is achieved by $U_\star$.}
\label{graph1}
\end{figure}

In Figure~\ref{graph1}, we
compare the 
error exponent for constant composition codes, achieved by
$U_\star$, to those associated with $U_{\mbox{\tiny MAP}}$ and $U_{\mbox{\tiny
ML}}$. The example considered refers to a quantized Gaussian channel with input distribution
\begin{align}
P_X(-3)=P_X(3)=0.05,\quad P_X(-1)=P_X(1)=0.45.
\end{align}
The channel output is quantized to 4 levels and the noise variance
is chosen such that the mutual information between input $X$ and the
quantized output is $0.5$~bits. The equivalent channel matrix is
\begin{align}
W=\begin{bmatrix} 0.8036  &  0.1964 &   0.0052 &   0.0000\\
    0.1912  &  0.6072 &   0.1912 &   0.0052\\
    0.0052  &  0.1912  &  0.6072 &   0.1912\\
    0.0000 &   0.0052 &   0.1964 &   0.8036
    \end{bmatrix}
\end{align}
where the $i$th column is a distribution on the output alphabet, given that the $i$th input symbol was transmitted.
 
We note that the gaps are considerable.

\subsection{The MAP Metric Achieves $I(X;Y)$}

We argue that a necessary and sufficient condition for a metric
$\{U(x|y),~x\in\calX,~y\in\calY\}$ to achieve the mutual information rate,
$I(X;Y)$ (induced by $P_X$ and $W$), is $U(x|y)=P_{X|Y}(x|y)$ (or an equivalent
metric), where $P_{X|Y}$
is the posterior distribution induced by $P_X\times W$. Indeed, let
$R=I(X;Y)-\epsilon$ be given, where $\epsilon>0$ is arbitrarily small.
Then, for $E_{\mbox{\tiny r}}(I(X;Y)-\epsilon,U)$ to be strictly positive,
there must exist $\rho\in[0,1]$ and $\theta\ge 0$ such that
\begin{equation}
-\sum_xP_X(x)\log\left(\sum_y\frac{W(y|x)}{[U_\theta(x|y)]^\rho}\right)+\rho[H(X|Y)+\epsilon]
> 0.
\end{equation}
Let $\theta> 0$ be such that there exists $\rho\in[0,1]$ for which this
condition holds true.
Now, for this given $\theta$, consider the function,
\begin{equation}
F(\rho)\dfn-\sum_xP_X(x)\log\left(\sum_y\frac{W(y|x)}{[U_\theta(x|y)]^\rho}\right)+\rho[H(X|Y)+\epsilon].
\end{equation}
It is easy to see that $F(0)=0$ and that $F(\rho)$ is concave, which means
that the derivative, $F'(\rho)$, is monotonically non-increasing. Therefore a
necessary and sufficient condition for the existence of $\rho\in[0,1]$ such
that $F(\rho) > 0$ (i.e., $\max_{\rho\in[0,1]}F(\rho)> 0$) is that $F'(0)> 0$.
Now,
\begin{eqnarray}
F'(0)&=&-\sum_xP_X(x)\cdot\frac{\sum_yW(y|x)[U_\theta(x|y)]^{-0}\log[1/U_\theta(x|y)]}{\sum_{y}W(y|x)[U_\theta(x|y)]^{-0}}
+H(X|Y)+\epsilon\\
&=&\sum_xP_X(x)\sum_yW(y|x)\log U_\theta(x|y)+H(X|Y)+\epsilon\\
&=&-D(P_{X|Y}\|U_\theta|P_Y)+\epsilon.
\end{eqnarray}
By the arbitrariness of $\epsilon> 0$, it follows that in order that $F'(0)> 0$ for all
$\epsilon>0$, the divergence term must vanish, and so, we must have that
$U_\theta(x|y)=P_{X|Y}(x|y)$ for all $x\in\calX$ and for all $y$ such that $P_Y(y)>0$.

Note that the optimal metric $U_*$ agrees with the MAP metric when $R=I(X;Y)$,
which corresponds to $\rho\to 0$, because in this case, the power $1/(1+\rho)$
tends to unity and $Z_\rho(x)$ tends to 1 for all $x$.

\section{Proof of Theorem 1}
\label{proof}

Beginning from the first part of the theorem,
let $\bx$ and $\by$ be the transmitted codeword and the received channel
output vector, respectively. Let $P_{\mbox{\tiny e}}(\bx,\by)$ denote the
expected error probability given that $\bx$ was transmitted and $\by$ was
received, where the expectation is with respect to the randomness of all other
codewords in $\calC$. Then,
\begin{eqnarray}
	\label{unionbound}
\bar{P}_{\mbox{\tiny
e}}(\bx,\by)&=&\mbox{Pr}\left[\bigcup_{\bx'\in\calC\setminus\{\bx\}}\left\{U(\bx',\by)\ge
U(\bx,\by)\right\}\right]\nonumber\\
&\le&\min\left\{1,2^{k-1}\sum_{\{\bx':~U(\bx',\by)\ge
U(\bx,\by)\}}2^{-nm}\right\}\nonumber\\
&\le&\min\left\{1,|\calM(\bx,\by)|2^{k-nm}\right\},
\end{eqnarray}
where
\begin{equation}
\calM(\bx,\by)=\{\bx':~U(\bx',\by)\ge U(\bx,\by)\},
\end{equation}
and where we have used the (truncated) union bound, the union being taken over
all $2^{k-1}$ pairwise error events, where an incorrect codeword $\bx'$,
randomly drawn under the uniform distribution over $\{0,1\}^{nm}$, happens to
have a metric score, $U(\bx',\by)$, that exceeds the one of the transmitted codeword,
$U(\bx,\by)$. In Appendix A, we show that this truncated union bound is
exponentially tight in spite of the fact that the codewords of a random linear
code are not mutually independent. This is done by deriving a lower bound of
the same exponential order.

Since $U(\bx',\by)$ depends on $(\bx',\by)$ only via their joint type, we can
assess the cardinality of $M(\bx,\by)$ by the method of types \cite{CK11}
as
\begin{eqnarray}
|\calM(\bx,\by)|&=&\sum_{\{\calT(\bx'|\by):~U(\bx',\by)\ge
U(\bx,\by)\}}|\calT(\bx'|\by)|\nonumber\\
&\exe&\sum_{\{\calT(\bx'|\by):~\log U(\bx',\by)\ge
\log U(\bx,\by)\}}\exp_2\left\{n\hat{H}_{\bx'\by}(X'|Y)\right\}\nonumber\\
&=&\exp_2\left\{n\max_{Q_{X'|Y}\in\calE(Q_{XY})}
H_Q(X'|Y)\right\}.
\end{eqnarray}
where
\begin{equation}
\calE(Q_{XY})=\left\{Q_{X'|Y}:~\sum_{x,y}Q_{X'Y}(x,y)\log U(x,y)\ge
\sum_{x,y}Q_{XY}(x,y)\log U(x,y)\right\}.
\end{equation}
It follows that
\begin{eqnarray}
\bar{P}_{\mbox{\tiny
e}}(\bx,\by)&\exe&\min\left\{1,\exp_2\left[k-nm+n\cdot\max_{Q_{X'|Y}\in\calE(Q_{XY})}
H_Q(X'|Y)\right]\right\}\nonumber\\
&=&\exp_2\left\{-n\left[m(1-\rfec)-\max_{Q_{X'|Y}\in\calE(Q_{XY})}
H_Q(X'|Y)\right]_+\right\}\nonumber\\
&=&\exp_2\left\{-n\left[H(P_X)-R-\max_{Q_{X'|Y}\in\calE(Q_{XY})}
H_Q(X'|Y)\right]_+\right\},
\end{eqnarray}
where in the last equality we have used the relation (\ref{rates}).
Averaging over the randomness of $\bY$, we get
\begin{eqnarray}
\bar{P}_{\mbox{\tiny
e}}(\bx)&=&\sum_{\by\in\calY^n}W(\by|\bx)P_{\mbox{\tiny
e}}(\bx,\by)\nonumber\\
&\exe&\sum_{\calT(\by|\bx)}|\calT(\by|\bx)|\cdot
\exp_2\left\{-n\left[H(P_X)-R-\max_{Q_{X'|Y}\in\calE(Q_{XY})}
H_Q(X'|Y)\right]_+\right\}\nonumber\\
&\exe&
\exp_2\bigg\{-n\min_{Q_{Y|X}}\bigg(D(Q_{Y|X}\|W|P_X)+\nonumber\\
& &\bigg[H(P_X)-R-\max_{Q_{X'|Y}\in\calE(Q_{XY})}
H_Q(X'|Y)\bigg]_+\bigg)\bigg\},
\end{eqnarray}
and since this expression depends on $\bx$ only via its type class
$\hat{P}_{\bx}=P_X$, then the same formula holds also for the average error
probability, which includes also the expectation w.r.t.\ the randomness of the
transmitted codeword, $\bx$, i.e.,
\begin{eqnarray}
\bar{P}_{\mbox{\tiny e}}&\exe&
\exp_2\bigg\{-n\min_{Q_{Y|X}}\bigg(D(Q_{Y|X}\|W|P_X)+\nonumber\\
& &\bigg[H(P_X)-R-\max_{Q_{X'|Y}\in\calE(Q_{XY})}
H_Q(X'|Y)\bigg]_+\bigg)\bigg\}.
\end{eqnarray}
It should be pointed out that this expression of the average error probability
is very similar to the one obtained for the ensemble of non-linear constant
composition codes for a given decoding metric, $U$. There is one important
difference, however: In the case of non-linear fixed composition codes, the
definition of the set $\calE(Q_{XY})$ should include the additional constraint that
$\sum_{y\in\calY}Q_Y(y)Q_{X'|Y}(x|y)=P_X(x)$ for 
every $x\in\calX$, whereas here, this constraint is absent.

We next derive the Lagrange-dual to this expression. Beginning from the inner
maximization, we have
\begin{eqnarray}
& &\max_{Q_{X'|Y}\in\calE(Q_{XY})}
H_Q(X'|Y)\nonumber\\
&=&\max_{Q_{X'|Y}}\inf_{\theta\ge
0}\bigg\{H_Q(X'|Y)+
\theta\sum_yQ_Y(y)\bigg[\sum_xQ_{X'|Y}(x|y)\log U(x,y)-\nonumber\\
& &\sum_xQ_{X|Y}(x|y)\log
U(x,y)\bigg]\bigg\}.
\end{eqnarray}
Moving on to the outer minimization, we obtain
\begin{eqnarray}
E_{\mbox{\tiny r}}(R,U)&=&\min_{Q_{Y|X}}\left(D(Q_{Y|X}\|W|P_X)+\bigg[H(P_X)-R-\max_{Q_{X'|Y}\in\calE(Q_{XY})}
H_Q(X'|Y)\bigg]_+\right)\nonumber\\
&=&\min_{Q_{Y|X}}\max_{0\le\rho\le 1}\left(D(Q_{Y|X}\|W|P_X)+
\rho\bigg[H(P_X)-R-\max_{Q_{X'|Y}\in\calE(Q_{XY})}
H_Q(X'|Y)\bigg]\right)\nonumber\\
&=&\min_{Q_{Y|X}}\max_{0\le\rho\le 1}\bigg(D(Q_{Y|X}\|W|P_X)+
\rho\bigg[H(P_X)-R-
\max_{Q_{X'|Y}}\inf_{\theta\ge
0}\bigg\{H_Q(X'|Y)+\nonumber\\
& &\theta\sum_yQ_Y(y)\bigg[\sum_xQ_{X'|Y}(x|y)\log
U(x,y)-\sum_xQ_{X|Y}(x|y)\log
U(x,y)\bigg]\bigg\}\bigg]\bigg)\nonumber\\
&=&\min_{Q_{Y|X}}\max_{0\le\rho\le 1}\min_{Q_{X'|Y}}\sup_{\theta\ge 0}\bigg(D(Q_{Y|X}\|W|P_X)+
\rho[H(P_X)-R]-
\rho H_Q(X'|Y)-\nonumber\\
& &\rho\theta\sum_yQ_Y(y)\bigg[\sum_xQ_{X'|Y}(x|y)\log
U(x,y)-\sum_xQ_{X|Y}(x|y)\log
U(x,y)\bigg]\bigg)\nonumber\\
&\eqa&\min_{Q_{Y|X}}\max_{0\le\rho\le 1}\min_{Q_{X'|Y}}\sup_{\theta\ge 0}\bigg(D(Q_{Y|X}\|W|P_X)+
\rho[H(P_X)-R]-
\rho H_Q(X'|Y)-\nonumber\\
	& &\hat{\theta}\sum_yQ_Y(y)\bigg[\sum_xQ_{X'|Y}(x|y)\log
U(x,y)-\sum_xQ_{X|Y}(x|y)\log
U(x,y)\bigg]\bigg)\nonumber\\
	&\eqb&\min_{Q_{Y|X}}\max_{0\le\rho\le 1}\sup_{\hat{\theta}\ge 0}\min_{Q_{X'|Y}}\bigg(D(Q_{Y|X}\|W|P_X)+
\rho[H(P_X)-R]-
\rho H_Q(X'|Y)-\nonumber\\
	& &\hat{\theta}\sum_yQ_Y(y)\bigg[\sum_xQ_{X'|Y}(x|y)\log
U(x,y)-\sum_xQ_{X|Y}(x|y)\log
U(x,y)\bigg]\bigg),
\end{eqnarray}
where in (a) we have defined $\hat{\theta}=\rho\theta$ 
and in (b) we have used the fact that the objective is convex in
$Q_{X'|Y}$ and affine in $\theta$.
Since the objective function is affine in $(\rho,\hat{\theta})$, then
after inner-most minimization over $Q_{X'|Y}$ it becomes concave in
$(\rho,\hat{\theta})$. The inner most minimization amounts to the maximization 
\begin{eqnarray}
& &\max_{Q_{X'|Y}}\left\{\rho
	H_Q(X'|Y)+\hat{\theta}\sum_yQ_Y(y)\sum_xQ_{X'|Y}(x|y)\log
U(x,y)\right\}\nonumber\\
&=&\rho\max_{Q_{X'|Y}}\sum_{y}Q_Y(y)\sum_{x}Q_{X'|Y}(x|y)
	\log\frac{[U(x,y)]^{\hat{\theta}/\rho}}{Q_{X'|Y}(x|y)}\nonumber\\
	&=&\rho\cdot\sum_yQ_Y(y)\log\left(\sum_x[U(x,y)]^{\hat{\theta}/\rho}\right)\nonumber\\
	&\dfn&\rho\cdot\sum_yQ_Y(y)\log Z(y,\hat{\theta}/\rho).
\end{eqnarray}
On substituting this back into the expression of $E_{\mbox{\tiny r}}(R,U)$, we
have
\begin{eqnarray}
E_{\mbox{\tiny r}}(R,U)&=&
	\min_{Q_{Y|X}}\max_{0\le\rho\le 1}\sup_{\hat{\theta}\ge
0}\bigg(D(Q_{Y|X}\|W|P_X)+\rho[H(P_X)-R]-\nonumber\\
	& &\rho\cdot\sum_{x,y}P_X(x)Q_{Y|X}(y|x)\log Z(y,\hat{\theta}/\rho)
	+\hat{\theta}\sum_xP_X(x)Q_{Y|X}(y|x)\log U(x,y)\bigg)\nonumber\\
	&\eqa&\max_{0\le\rho\le 1}\sup_{\hat{\theta}\ge
0}\min_{Q_{Y|X}}\bigg(\sum_{x,y}P_X(x)Q_{Y|X}(y|x)\log\frac{Q_{Y|X}(y|x)}{W(y|x)}
+\rho[H(P_X)-R]-\nonumber\\
	& &\rho\cdot\sum_{x,y}P_X(x)Q_{Y|X}(y|x)\log Z(y,\hat{\theta}/\rho)
	+\hat{\theta}\sum_xP_X(x)Q_{Y|X}(y|x)\log U(x,y)\bigg)\nonumber\\
	&=&\max_{0\le\rho\le 1}\sup_{\hat{\theta}\ge
0}\min_{Q_{Y|X}}\bigg(\sum_{x,y}
	P_X(x)Q_{Y|X}(y|x)\log\frac{Q_{Y|X}(y|x)[U(x,y)]^{\hat\theta}}{W(y|x)[Z(y,\hat{\theta}/\rho)]^\rho}
+\rho[H(P_X)-R]\bigg)\nonumber\\
	&=&\max_{0\le\rho\le 1}\sup_{\hat{\theta}\ge
0}\bigg(-\sum_x
	P_X(x)\log\left(\sum_y\frac{W(y|x)Z(y,\hat{\theta}/\rho)]^\rho}{[U(x,y)]^{\hat{\theta}}}\right)
+\rho[H(P_X)-R]\bigg)\nonumber\\
	&\eqb&\max_{0\le\rho\le 1}\sup_{\theta\ge
	0}\bigg[-\sum_x
	P_X(x)\log\left(\sum_y\frac{W(y|x)}{[U_\theta(x|y)]^{\rho}}\right)
	+\rho[H(P_X)-R]\bigg],
\end{eqnarray}
where in (a) we have used the fact that the objective is convex 
in $Q_{X'|Y}$ and concave in $(\rho,\hat{\theta})$, and in (b) we returned
to the original optimization parameter, $\theta=\hat{\theta}/\rho$.
This completes the proof of part 1 of the theorem.

Moving on to part 2 of the theorem,
in Appendix B, we prove that the following
(Lagrange-dual) expression may serve as an
alternative representation of $E_{\mbox{\tiny r}}^{\mbox{\tiny cc}}(R)$:
\begin{equation}
\label{ccld}
E_{\mbox{\tiny r}}^{\mbox{\tiny cc}}(R)=\min_V\max_{0\le\rho\le
1}\bigg\{-(1+\rho)\sum_{x}P_X(x)\log\left[
\sum_y\left(W(y|x)[V(y)]^\rho\right)^{1/(1+\rho)}\right]-\rho R\bigg\},
\end{equation}
where the minimization is over all probability assignments,
$V=\{V(y)\}$ over $\calY$.

Next, consider the following chain of equalities and inequalities:
\begin{eqnarray}
\label{chain}
& &\sum_xP_X(x)\log\left(\sum_y\frac{W(y|x)}{[U_\theta(x|y)]^\rho}\right)\nonumber\\&=&
\sum_xP_X(x)\log\left(\sum_y\tW(y|x)\frac{W(y|x)}{[U_\theta(x|y)]^\rho
\tW(y|x)}\right)\nonumber\\
&\gea&\max_{\tW}\sum_xP_X(x)\sum_y\tW(y|x)\log\left(\frac{W(y|x)}{[U_\theta(x|y)]^\rho
\tW(y|x)}\right)\nonumber\\
&=&\max_{\tW}\left\{\sum_xP_X(x)\sum_y\tW(y|x)\log\left(\frac{W(y|x)}{\tW(y|x)}\right)-\rho\sum_{x,y}P_X(x)\tW(y|x)\log
U_\theta(x|y)\right\}\nonumber\\
&\geb&\max_{\tW}\left\{\sum_xP_X(x)\sum_y\tW(y|x)\log\left(\frac{W(y|x)}{\tW(y|x)}\right)+\rho
\tH(X|Y)\right\}\nonumber\\
&=&\max_{\tW}\left\{\sum_xP_X(x)\sum_y\tW(y|x)\log\left(\frac{W(y|x)}{\tW(y|x)}\right)+\rho
[H(X)+\tH(Y|X)-\tH(Y)]\right\}\nonumber\\
&\eqc&\max_{\tW}\max_V\sum_xP_X(x)\sum_y\tW(y|x)\bigg[\log\left(\frac{W(y|x)}{\tW(y|x)}\right)+\nonumber\\
& &\rho H(X)-\rho\log \tW(y|x)+\rho\log V(y)\bigg]\nonumber\\
&=&\max_V\max_{\tW}\sum_xP_X(x)\sum_y\tW(y|x)\left[\log\left(\frac{W(y|x)[V(y)]^\rho}{[\tW(y|x)]^{1+\rho}}\right)+\rho
H(X)\right]\nonumber\\
&=&\max_V\max_{\tW}\sum_xP_X(x)\sum_y\tW(y|x)\left[(1+\rho)\log\left(\frac{(W(y|x)[V(y)]^\rho)^{1/(1+\rho)}}{\tW(y|x)}\right)+\rho
H(X)\right]\nonumber\\
&=&\max_V\left\{(1+\rho)\sum_xP_X(x)\log\left(\sum_y(W(y|x)[V(y)]^\rho)^{1/(1+\rho)}\right)+\rho
H(X)\right\}
\end{eqnarray}
where in (a) we have used Jensen's inequality, 
in (b) and onward, $\tH(Y|X)$ and $\tH(Y)$ refer to entropies induced by
$P_X\times\tW$, and in (c), $V$ is a probability distribution on $\calY$.
Consequently,
\begin{eqnarray}
& &-\sum_xP_X(x)\log\left(\sum_y\frac{W(y|x)}{[U_\theta(x|y)]^\rho}\right)+\rho[H(X)-R]\nonumber\\
&\le&\min_V\left\{-(1+\rho)\sum_xP_X(x)\log\left(\sum_y(W(y|x)[V(y)]^\rho)^{1/(1+\rho)}\right)-\rho
H(X)+\rho[H(X)-R)\right\}\nonumber\\
&=&\min_V\left\{-(1+\rho)\sum_xP_X(x)\log\left(\sum_y(W(y|x)[V(y)]^\rho)^{1/(1+\rho)}\right)-\rho
R\right\}
\end{eqnarray}
and after maximizing both sides over $\rho\in[0,1]$ and $\theta\ge 0$, we get
$E_{\mbox{\tiny r}}(R,U)\le E_{\mbox{\tiny r}}^{\mbox{\tiny cc}}(R)$, in view
of eq.\ (\ref{ccld}). This completes the proof of part 2 of the theorem.

To prove part 3 of the theorem,
let us examine the conditions under which the inequalities in (\ref{chain}) become equalities:
The first inequality, which is an application of Jensen's inequality, is met
with equality if the expression,
$$\frac{W(y|x)}{[U_\theta(x|y)]^\rho
\tW(y|x)}$$ 
is independent of $\by$, though it is still allowed to depend on $x$.
In other words, $U_\theta(x|y)$ must satisfy
\begin{equation}
\frac{W(y|x)}{\tW(y|x)U_\theta(x|y)^\rho}= K(x),
\end{equation}
for some function $K(x)$.
The second inequality becomes equality if
\begin{equation}
U_\theta(x|y)= \frac{P_X(x)\tW(y|x)}{\sum_{x'}P(x')\tW(y|x')}.
\end{equation}
Now, the optimal $\tW$, that achieves the last line of (\ref{chain}), is given by
\begin{equation}
\tW(y|x)=\frac{[W(y|x)]^{1/(1+\rho)}[V(y)]^{\rho/(1+\rho)}}{\sum_{y'}[W(y'|x)]^{1/(1+\rho)}[V(y')]^{\rho/(1+\rho)}}
\dfn\frac{[W(y|x)]^{1/(1+\rho)}[V(y)]^{\rho/(1+\rho)}}{Z(x)}.
\end{equation}
We argue that the following metric satisfies both requirements.
\begin{equation}
U_\theta(x|y)=\frac{P_X(x)[W(y|x)]^{1/(1+\rho)}/Z(x)}
{\sum_{x'}P_X(x')[W(y|x')]^{1/(1+\rho)}/Z(x')}\dfn\frac{P_X(x)[W(y|x)]^{1/(1+\rho)}/Z(x)}{\zeta(y)}.
\end{equation}
Indeed,
\begin{eqnarray}
\label{1st}
& &\frac{W(y|x)}{\tW(y|x)[U_\theta(x|y)]^\rho}\nonumber\\
&=&\frac{W(y|x)Z(x)\zeta^\rho(y)Z^\rho(x)}{[W(y|x)]^{1/(1+\rho)}[V(y)]^{\rho/(1+\rho)}P_X^\rho(x)[W(y|x)]^{\rho/(1+\rho)}}\\
&=&\frac{[Z(x)]^{1+\rho}\zeta^\rho(y)}{[V(y)]^{\rho/(1+\rho)}P_X^\rho(x)}
\end{eqnarray}
Now, observe that the maximizing $V$ in (\ref{chain}) is given by
\begin{equation}
V(y)=\sum_xP_X(x)\tW(y|x)=\sum_xP_X(x)\cdot\frac{[W(y|x)]^{1/(1+\rho)}[V(y)]^{\rho/(1+\rho)}}{Z(x)}
\end{equation}
or, equivalently, by dividing both sides by $[V(y)]^{\rho/(1+\rho)}$,
\begin{equation}
[V(y)]^{1/(1+\rho)}=\sum_xP_X(x)\cdot\frac{[W(y|x)]^{1/(1+\rho)}}{Z(x)},
\end{equation}
and so, raising both sides to the power of $\rho$, we get
\begin{equation}
[V(y)]^{\rho/(1+\rho)}=\left(\sum_xP_X(x)\cdot\frac{[W(y|x)]^{1/(1+\rho)}}{Z(x)}\right)^\rho=\zeta^\rho(y),
\end{equation}
so the ratio $\zeta^\rho(y)/[V(y)]^{\rho/(1+\rho)}=1$, which, together with
(\ref{1st}), implies that 
\begin{equation}
\frac{W(y|x)}{\tW(y|x)[U_\theta(x|y)]^\rho}=\frac{[Z(x)]^{1+\rho}}{P_X^\rho(x)},
\end{equation}
which is independent of $y$, as required.

As for the second requirement,
\begin{eqnarray}
U_\theta(x|y)&=&\frac{P_X(x)\tW(y|x)}{\sum_{x'}P_X(x')\tW(y|x')}\nonumber\\
&=&\frac{P_X(x)[W(y|x)]^{1/(1+\rho)}[V(y)]^{\rho/(1+\rho)}/Z(x)}{\sum_{x'}P_X(x')[W(y|x')]^{1/(1+\rho)}
[V(y)]^{\rho/(1+\rho)}/Z(x')}\nonumber\\
&=&\frac{P_X(x)[W(y|x)]^{1/(1+\rho)}/Z(x)}{\sum_{x'}P_X(x')[W(y|x')]^{1/(1+\rho)}
/Z(x')}.
\end{eqnarray}
Finally, the relationship between $Z(x)$ and $V(y)$ is as follows: on the one hand,
by definition,
\begin{equation}
Z(x)=\sum_y[W(y|x)]^{1/(1+\rho)}[V(y)]^{\rho/(1+\rho)},
\end{equation}
and on the other hand, we saw that
\begin{equation}
V(y)=\left[\sum_x\frac{P_X(x)[W(y|x)]^{1/(1+\rho)}}{Z(x)}\right]^{1+\rho}.
\end{equation}
On substituting the second relation into the first one, we end up with
following system of equations in the vector $Z=\{Z(x),~x\in\calX\}$:
\begin{equation}
Z(x)=\sum_y[W(y|x)]^{1/(1+\rho)}\left[\sum_{x'}\frac{P_X(x')[W(y|x')]^{1/(1+\rho)}}{Z(x')}\right]^\rho,~~~~\forall~x\in\calX
\end{equation}
In summary, assuming that there exists a solution to this set of equations, the
metric 
\begin{equation}
U_\star(x|y,\rho)=\frac{P(x)[W(y|x]^{1/(1+\rho)}/Z(x)}{\sum_{x'}P(x')[W(y|x')]^{1/(1+\rho)}/Z(x')}.
\end{equation}
saturates all inequalities in (\ref{chain}) at the same time, and hence, after
optimization over $\rho$, achieves $E_{\mbox{\tiny r}}^{\mbox{\tiny cc}}(R)$.
\section{Conclusions}

In this work, we have shown that the code mismatch of using for a constant composition code a linear extension code for decoding can be fully compensated by using a mismatched additive decoding metric. That is, the error exponent \cite[Theorem~10.2]{CK11} is achieved and in particular, the capacity of any DMC can be achieved by decoding a linear code with an additive MAP decoding metric.

Interesting direction for future research are:
\begin{enumerate}
\item In this work, we have considered DMCs and we used the method of types in our theoretical derivations. Can the presented results be generalized to continuous output channels?
\item For finite length, constant composition codes may not be optimal and minimum cost DM~\cite{schulte2019divergence} may be preferable. A theoretic analysis may find a replacement of constant composition codes that is optimal in the finite length regime.
\item In this work, decoding was our main focus. Future work may include practical encoding aspects, for instance, by assuming the PAS~\cite{BSS15} architecture or the LLPS~\cite{BLCS19} generalization.
\end{enumerate}

\section*{Appendix A}
\renewcommand{\theequation}{A.\arabic{equation}}
    \setcounter{equation}{0}

In this appendix, we show that the first inequality in eq.\ (\ref{unionbound})
is exponentially tight.
This is done by deriving a matching lower bound of the same exponential order
as the upper bound.

Owing to the results of Domb, Zamir and Feder \cite{DZF16}, we begin with the
following observation. Consider three distinct messages $w$, $w'$, and $\tilde{w}$.
First, if their binary representations, $\bb(w)$, $\bb(w')$, and
$\bb(\tilde{w})$ are linearly independent,
the three corresponding codewords are statistically mutually independent by
the independently sampled rows of
$\bG$. If the binary representations are
linearly dependent, i.e., $\bb(\tilde{w)}=\bb(w)\oplus\bb(w')$, then let us
define
\begin{align}
        \ba(w)=\bb(w)\cdot\bG,\quad \ba(w')=\bb(w')\cdot\bG,
\end{align}
and observe that $\ba(w)$, $\ba(w')$, and $\bv$ are statistically mutually
independent.
The corresponding codewords are given by
\begin{eqnarray}
        \bc(w)&=&\ba(w)\oplus\bv\\
        \bc(w')&=&\ba(w')\oplus\bv\\
        \bc(\tilde{w})&=&\ba(w)\oplus\ba(w')\oplus\bv,
\end{eqnarray}
and therefore, the inverse transformation is given by
\begin{eqnarray}
        \ba(w)&=&\bc(w')\oplus\bc(\tilde{w})\\
        \ba(w')&=&\bc(w)\oplus\bc(\tilde{w})\\
        \bv&=&\bc(w)\oplus\bc(w')\oplus\bc(\tilde{w}).
\end{eqnarray}
Since the transformation between these two triples of vectors is one-to-one,
and every triple $(\ba(w),\ba(w'),\bv)$ has probability $2^{-3nm}$,
then the same is true for every triple $(\bc(w),\bc(w'),\bc(\tilde{w}))$,
which means that the three codewords are mutually independent and each one of
them is
uniformly distributed across $\{0,1\}^{nm}$.

Consider the pairwise error events,
\begin{align}
        A_{w'}=\left\{U(\bx(w'),\by)\geq U(\bx(w),\by)\right\}.
\end{align}
Since the codewords are pairwise independent, message $w'$
is assigned to any particular codeword with
probability $2^{-nm}$ and the probability of event $A_{w'}$ is
\begin{align}
        \Pr(A_{w'})=|\calM(\bx(w),\by)|\cdot 2^{-nm}\dfn \alpha.
\end{align}
Since the codewords are triple-wise
independent, we can use de Caen's lower bound \cite{deCaen97}
to the probability of a union, and obtain
\begin{align}
\bar{P}_{\mbox{\tiny e}}(\bx(w),\by)&=\Pr\Bigl(\bigcup_{w'\neq
w}A_{w'}\Bigr)\\
        &\geq \sum_{w'\neq w}\frac{[\Pr(A_{w'})]^2}{\sum_{\tilde{w}\neq
w}\Pr(A_{w'}\cap A_{\tilde{w}})}\\
        &= \sum_{w'\neq
w}\frac{[\Pr(A_{w'})]^2}{\Pr(A_{w'})+\sum_{\tilde{w}\neq w,w'}\Pr(A_{w'}\cap
A_{\tilde{w}})}\\
&= \sum_{w'\neq w}\frac{\alpha^2}{\alpha+\sum_{\tilde{w}\neq
w,w'}\Pr(A_{w'}\cap A_{\tilde{w}})}.\label{eq:bound with fraction}
\end{align}
Next, we consider the term $\Pr(A_{w'}\cap A_{\tilde{w}})$. Here, the three
channel codewords
$\bx(w)$, $\bx(w')$, and $\bx(\tilde{w})$ are involved, with the three
messages
being pairwise distinct. By the triple-wise independence, $\bx(w')$ and
$\bx(\tilde{w})$ are conditionally independent given $\bx(w)$,
and so,
\begin{align}
\Pr(A_{w'}\cap A_{\tilde{w}})=\Pr(A_{w'})\cdot\Pr(A_{\tilde{w}})=\alpha^2.
\end{align}
Continuing with \eqref{eq:bound with fraction}, we have
\begin{align}
        \bar{P}_{\mbox{\tiny e}}(\bx(w), \by)&\geq \sum_{w'\neq
w}\frac{\alpha^2}{\alpha+(2^k - 2)\alpha^2}\\
&=\frac{(2^k-1)\alpha}{1+(2^k - 2)\alpha}\\
&\geq\frac{(2^k-1)\alpha}{2\cdot\max\left\{ (2^k - 2)\alpha, 1\right\}}\\
&\geq\frac{(2^k-2)\alpha}{2\cdot\max\left\{ (2^k - 2)\alpha, 1\right\}}\\
&=\min\left\{\frac{1}{2}, \frac{(2^k- 2)\alpha}{2}\right\}\\
        &\doteq \min\left\{1,|\calM(\bx(w),\by)|2^k
2^{-nm}\right\}\label{eq:linear upper bound},
\end{align}
which is of the same exponential order as the truncated union bound in
(\ref{unionbound}).

\section*{Appendix B}
\renewcommand{\theequation}{B.\arabic{equation}}
    \setcounter{equation}{0}

In this appendix, we prove the alternative form of the random coding error exponent for constant composition codes.
\begin{eqnarray}
& &\min_{Q_{Y|X}}\left\{D(Q_{Y|X}\|W|P_X)+[I_Q(X;Y)-R]_+\right\}\nonumber\\
&=&\min_{Q_{Y|X}}\max_{0\le\rho\le
1}\left\{D(Q_{Y|X}\|W|P_X)+\rho[H_Q(Y)-H_Q(Y|X)-R]\right\}\nonumber\\
&=&\min_{Q_{Y|X}}\max_{0\le\rho\le 1}\bigg\{\sum_{x,y}P_X(x)Q_{Y|X}(y|x)
	\bigg[\log\frac{Q_{Y|X}(y|x)}{W(y|x)}+\rho\log\frac{1}{Q_Y(y)}+\nonumber\\
	& &\rho\log Q_{Y|X}(y|x)\bigg]-\rho R\bigg\}\nonumber\\
&\eqa&\min_{Q_{Y|X}}\max_{0\le\rho\le 1}\min_V\bigg\{\sum_{x,y}P_X(x)Q_{Y|X}(y|x)
	\bigg[\log\frac{Q_{Y|X}(y|x)}{W(y|x)}+\rho\log\frac{1}{V(y)}+\nonumber\\
	& &\rho\log Q_{Y|X}(y|x)\bigg]-\rho R\bigg\}\nonumber\\
&\eqb&\min_{Q_{Y|X}}\min_V\max_{0\le\rho\le 1}\bigg\{\sum_{x,y}P_X(x)Q_{Y|X}(y|x)
	\bigg[\log\frac{Q_{Y|X}(y|x)}{W(y|x)}+\rho\log\frac{1}{V(y)}+\nonumber\\
	& &\rho\log Q_{Y|X}(y|x))-\rho R\bigg\}\nonumber\\
&=&\min_V\min_{Q_{Y|X}}\max_{0\le\rho\le 1}\bigg\{\sum_{x,y}P_X(x)Q_{Y|X}(y|x)
	\bigg[\log\frac{Q_{Y|X}(y|x)}{W(y|x)}+\rho\log\frac{1}{V(y)}+\nonumber\\
	& &\rho\log Q_{Y|X}(y|x)\bigg]-\rho R\bigg\}\nonumber\\
&\eqc&\min_V\max_{0\le\rho\le 1}\min_{Q_{Y|X}}\left\{\sum_{x,y}P_X(x)Q_{Y|X}(y|x)
\log\frac{[Q_{Y|X}(y|x)]^{1+\rho}}{W(y|x)[V(y)]^\rho}-\rho R\right\}\nonumber\\
&=&\min_V\max_{0\le\rho\le 1}\min_{Q_{Y|X}}\left\{(1+\rho)\sum_{x,y}P_X(x)Q_{Y|X}(y|x)
\log\frac{Q_{Y|X}(y|x)}{(W(y|x)[V(y)]^\rho)^{1/(1+\rho)}}-\rho R\right\}\nonumber\\
&=&\min_V\max_{0\le\rho\le 1}\left\{-(1+\rho)\sum_{x}P_X(x)\log\left[
\sum_y(W(y|x)[V(y)]^\rho)^{1/(1+\rho)}\right]-\rho R\right\}\nonumber\\
&=&\min_V\max_{0\le\rho\le 1}\bigg\{-\sum_{x}P_X(x)\log\left[
	\sum_y\left(\frac{W(y|x)[V(y)]^\rho}{[P_X(x)]^\rho}\right)^{1/(1+\rho)}\right]^{1+\rho}+\nonumber\\
& &\rho[H(P_X)-R]\bigg\},
\end{eqnarray}
where the inner-most minimization in (a) is over all probability distributions $\{V(y)\}$ on $\calY$,
(b) holds because the objective is convex in $V$ and concave in $\rho$,
and similarly, (c) is because the objective is convex in $Q_{Y|X}$ and concave in $\rho$.

\end{document}